\documentclass[12pt]{article}
\usepackage[textures]{graphics}
\parskip=\bigskipamount
\newcommand{\beq}{\begin{equation}}
\newcommand{\eeq}{\end{equation}}
\newcommand{\beqa}{\begin{eqnarray}}
\newcommand{\eeqa}{\end{eqnarray}}
\newcommand{\ket} [1] {\vert #1 \rangle}
\newcommand{\bra} [1] {\langle #1 \vert}
\newcommand{\braket}[2]{\langle #1 | #2 \rangle}

\begin{document}
\title{\bf If $1=2\oplus 3$, then $1=2\odot 3$: Bell states, finite groups,
 and mutually unbiased bases, a unifying approach.}
\date{}
\author{}
\maketitle
\vglue -1.8truecm
\centerline{\large Thomas Durt\footnote{TENA-TONA Free University of
Brussels, Pleinlaan 2, B-1050 Brussels, Belgium. email: thomdurt@vub.ac.be}}\bigskip\bigskip


{\it Abstract: We study the relationship between Bell states, finite groups and complete sets of bases. 
We show how to obtain a set of $N+1$ bases in which Bell states are invariant.
 They generalize the $X$, $Y$ and $Z$ qubit bases and are associated to groups of unitary transformations that generalize
  the sigma operators of Pauli. When the dimension $N$ is a prime power, we derive (in agreement with well-known results) 
  a set of mutually unbiased bases. We show how they can be expressed
   in terms of the (operations of the) associated finite field of $N$ elements. 
    }

\section*{Introduction}

We showed recently \cite{DurtNagler}, in the framework of quantum cloning, that in dimension $N$ = 4 different classes of Bell states can be defined, that are associated to different groups of permutations 
of the $N$ basis states. These Bell states were also shown to be invariant in the dual basis that can be associated to the permutation group. We show
 in the present paper that this construction can be generalized, 
and that we can generate a group of $N^2$ unitary transformations, 
that consists of $N+1+i$ subgroups with $N$ elements ($N-1$ elements plus the identity), which are in one to one correspondence
 with $N+1+i$ bases ($i$ is a positive integer equal to zero in prime dimensions only). We show that these bases form a complete set in the sense that they
  allow us to perform a full tomography of an arbitrary quantum state.
  
    We recover, in a synthetic formulation, the well-known results that can be found in the litterature relatively to the 
  existence of complete sets of mutually\footnote{Two orthonormal bases of a $N$ dimensional Hilbert space are said to be mutually 
  unbiased if whenever we choose one state in the first basis, and a second state in the second basis,
   the modulus squared of their in-product is equal to $1/N$. } unbiased bases \cite{Schwinger} in dimensions $p$ and $p^m$
   (with $p$ prime and $m$ a positive integer \cite{Wootters,Zeil}).
  A crucial element of the construction is the existence of a finite commutative division ring (field\footnote{A field is a set with a multiplication and an addition operation which satisfy the usual rules, associativity and commutativity of both operations, the distributive law, existence of an additive identity 0 and a multiplicative identity 1, additive inverses, and multiplicative inverses for every element, 0 excepted. })
   of $N$
   elements. As it is well known, finite fields with $N$ elements exist if and only if the
     dimension $N$ is a prime or a power of
     a prime, and the derivation of a set of mutually
      unbiased bases is already known in such cases. We derive in the last section an expression for the mutually unbiased bases, in terms of the underlying field, in prime power dimensions.
       Beside, our approach also provides a complete set of (not necessary mutually unbiased)
       bases in arbitrary dimensions. Actually, in the example that we work out in detail in the second section (for which the essential symmetry group is the cyclic shift modulo $N$),
        such bases are eigenbases of the groups of error operators
        that found many applications in quantum information. Nevertheless, our method is valid in more general situations 
        and can be applied directly to other symmetry groups. For instance, when the dimension is a prime or the power of a prime, we recover the well-known complete set of mutually unbiased bases,
         but the novelty is that they appear, in our approach,
         to be related to a finite group of symmetry, a group that maps
          Bell states onto Bell states, which in turn is intimately related to a subgroup of the permutation group of $N$ elements. 
          There exist several applications of these methods in the framework of quantum 
          information (the mean king's problem \cite{vaid,Englert,2003,arXiv}, error correction \cite{nielsen},
          quantum cryptography (\cite{BB84} to  \cite{DurtKwek}) and so on. 
          Our approach makes it possible to derive all the useful tools, in a self-consistent manner, once we identified the relevant symmetry
           of the problem under consideration, which simplifies considerably the treatment.

    \section{The dual basis}
Let us consider $C$, the generator of the cyclic
permutations that shifts each label of the states of the computational basis ($\{\ket{0},\ket{1},...,\ket{N-1}\}$) by unity 
($l \to l+1$ (${\rm mod} N$)). These permutations form a commutative 
group that is isomorphic to the addition modulo $N$. It is easy to check that the
 multiplication (modulo $N$) forms
  a commutative ring which is  distributive relatively to the addition. 
  generalizing the procedure outlined in \cite{DurtNagler}, 
  we define the dual basis as follows:

\beqa
\ket{\tilde j}_{F}=\frac{1}{\sqrt{N}}\sum_{k=0}^{N-1} \gamma_{F}^{(k\odot_{F} j)}\ket{k }\eeqa

In the equation 1, the symbol $\odot_{F}$ represents the multiplication modulo $N$,
 while $\gamma_{F}$ is taken to be equal to the $N$th root of unity 
 ($\gamma_{F} =e^{ i .2\pi/N}$). The 
 index $F$ is aimed at indicating that the dual basis is the discrete Fourier transform
  of the computational basis. For that choice of addition and multiplication, we have, of course,
  \beqa \ket{\tilde j}_{F}=\frac{1}{\sqrt{N}}\sum_{k=0}^{N-1}e^{2\pi i(k\cdot j) /N}\ket{k }\label{fourierdual}\eeqa 
  Remark that 
 this definition can be enlarged in order to cover the case of different additions and multiplications as we shall see soon. The general expression is: 
 \beqa \ket{\tilde j}_{G}=\frac{1}{\sqrt{N}}\sum_{k=0}^{N-1} \gamma_{G}^{(k\odot_{G} j)}\ket{k }\eeqa
 In the following, we shall usually omit the index $G$ for the states in order not to complicate the notations. 
It is easy to check that, in the Fourier case (modulo $N$ operations) the dual states are invariant, up to a global phase, under (the cyclic group generated by) $C$. Indeed, we have:
\beqa
C.\ket{\tilde j}=\frac{1}{\sqrt{N}}\sum_{k=0}^{N-1} e^{2\pi i(k\odot_{F} j) /N}\ket{k \oplus_{F} 1}\\
=\frac{1}{\sqrt{N}}\sum_{k=0}^{N-1}e^{2\pi i((k'-1)\cdot j) /N}\ket{k' }=e^{2\pi i((-1)\cdot j) /N}\ket{\tilde j}\\
=\gamma_{F}^{((\ominus_{F} 1)\odot_{F} j) }\ket{\tilde j}\eeqa where the symbols $\oplus_{F}$ and
 $\ominus_{F}$ represent the addition modulo $N$, and its inverse. 
 Beside, as we showed in refs. \cite{DurtKwek} and \cite{DurtNagler}, we can define the generalized Bell states\footnote{The use of Bell states in the theory of cloning machines was systematized by N. Cerf \cite{CERFPRL}. 
 We were led to introduce the present definition of (generalized) Bell states for practical reasons related to the invariance of the cloning state.} as follows:
 
 \beqa
\ket{B_{m^*,n}}=N^{-1/2} \sum_{k=0}^{N-1} \gamma^{(k\odot_{F} j)}
\ket{k^*}\ket{k+m}\\
=N^{-1/2} \sum_{k=0}^{N-1} e^{2\pi i (kn/N)} \ket{k^*}\ket{k+m}
\nonumber \label{bell}
\eeqa
In this definition, we introduced the basis $\ket{k^*}$ which is the complex conjugate basis
 of the direct basis $\ket{k}$. This does not make any difference when $\ket{k}$ is 
 the reference (computational) basis but it does when the Bell states are defined relatively to a basis that possesses states with complex amplitudes when they are expanded in the computational basis. For instance, in the dual basis, we find the following Bell states:
 \beqa
\ket{\tilde B_{m^*,n}}=N^{-1/2} \sum_{k=0}^{N-1} \gamma_{F}^{(k\odot_{F} j)}
\ket{\tilde k^*}\ket{\tilde{k\oplus_{F}m}}\\
=N^{-1/2} \sum_{k=0}^{N-1} e^{2\pi i (kn/N)} \ket{\tilde k^*}\ket{\tilde{k\oplus_{F}m}}\nonumber \\
=N^{-3/2} \sum_{k,l,j=0}^{N-1} e^{2\pi i (kn/N)}e^{2\pi i (-kl/N)}
e^{2\pi i (k+m)j/N)} \ket{l^*}\ket{j}\nonumber  \\
=N^{-3/2} N\sum_{l,j=0}^{N-1}\delta_{n-l+j} e^{2\pi i (mj/N)} 
\ket{l^*}\ket{j}\nonumber \\
=N^{-1/2} \sum_{l=0}^{N-1} e^{2\pi i (mj/N)} \ket{l^*}\ket{l\ominus_{F}n}=
\ket{\tilde B_{\ominus_{F}n^*,m}}
\label{bellinv}
 \eeqa
 We made use of the essential property 
 \beq\sum_{p=0}^{N-1} e^{2\pi i (p\cdot q/N)}=N\delta_{q,0},\eeq
  where $p$ and $q$ are integer numbers, and $N$ an arbitrary non-null integer number. 
  Note that this property can be generalized to: 
\beq\sum_{p=0}^{N-1} \gamma^{ (p\odot q)}=N\delta_{q,0}\eeq 

Indeed, the dual Fourier basis is not the unique interesting dual basis that can be associated to the computational basis. 
We showed for instance in Ref.\cite{DurtNagler} that, 
in dimension $N=4$, there exists another interesting dual basis (the double Hadamard transform of the 
computational basis) that can be re-expressed as follows:
\beqa
\ket{\tilde j}=\frac{1}{\sqrt{N}}\sum_{k=0}^{N-1} \gamma_{H}^{(k\odot_{H} j)}\ket{k }\\
=\frac{1}{\sqrt{N}}\sum_{k=0}^{N-1}(-1)^{ (k\odot_{H} j) }\ket{k }\eeqa

This basis corresponds to another choice of $\gamma$ (here $\gamma_{H}=-1$). The
 corresponding mutiplication, which is listed in the table 1 is also different from the multiplication modulo 4. The transformations matrices between the computational basis and the dual basis are given in tables \ref{tab3F} and \ref{tab3H}. 
 
\begin{table}
\begin{tabular}{c||c|c|c|c}
\hline $\odot_{H}$  & $0$ & $1$ & $2$  & $3$\\
\hline \hline  0 & 0 & $0$ & $0$  & $0$ \\
1 & $0$ & $1$ & $2$  & $3$\\
2 & $0$ & $2$ & $3$  & $1$\\
3 & $0$ & $3$ & $1$  & $2$ \\
 \hline
\end{tabular}
\caption{The Hadamard product in dimension 4. }\label{tab1}
\end{table}

The Hadamard multiplication in turn is distributive relatively to the addition listed in table 2.

\begin{table}
\begin{tabular}{c||c|c|c|c}
\hline $\oplus_{H}$  & $0$ & $1$ & $2$  & $3$\\
\hline \hline  0 & 0 & $1$ & $2$  & $3$ \\
1 & $1$ & $0$ & $3$  & $2$\\
2 & $2$ & $3$ & $0$  & $1$\\
3 & $3$ & $2$ & $1$  & $0$ \\
 \hline
\end{tabular}
\caption{The Hadamard addition in dimension 4. }\label{tab2}
\end{table}
Both operations are commutative as can be seen from the symmetry of the tables 1 and 2 under transposition. Moreover, the multiplication table, amputed from the first line and column exhibits an invertible  (group) structure.
\begin{table}
\begin{tabular}{c||c|c|c|c}
\hline $\odot_{F}$  & $0$ & $1$ & $2$  & $3$\\
\hline \hline  0 & 0 & $0$ & $0$  & $0$ \\
1 & $0$ & $1$ & $2$  & $3$\\
2 & $0$ & $2$ & $0$  & $2$\\
3 & $0$ & $3$ & $2$  & $1$ \\
 \hline
\end{tabular}
\caption{The Fourier product in dimension 4. }\label{tab3}
\end{table}
\begin{table}
\begin{tabular}{c||c|c|c|c}
\hline $\oplus_{F}$  & $0$ & $1$ & $2$  & $3$\\
\hline \hline  0 & 0 & $1$ & $2$  & $3$ \\
1 & $1$ & $2$ & $3$  & $0$\\
2 & $2$ & $3$ & $0$  & $1$\\
3 & $3$ & $0$ & $1$  & $2$ \\
 \hline
\end{tabular}
\caption{The Fourier addition in dimension 4. }\label{tab4}
\end{table}
Just to give an idea, the corresponding operations, in the Fourier case, 
(multiplication and addition modulo 4) are listed in tables 3 and 4. One can check that the multiplication modulo 4 is distributive relatively to the addition modulo 4.
In the Fourier case in dimension 4 however, the amputed table of multiplication does not represent a group because it contains the element 0,
 so that the amputed multiplication is not a closed operation. An important property, that characterizes $\gamma_{F}$ and $\gamma_{H}$ as well, 
 is that they represent the corresponding additive groups, more precisely they are endowed with the following property:
 $\gamma_G^{i\odot_G j}\gamma_G^{i\odot_G k}=\gamma^{i\odot_G (j \oplus_G k)}$,
  where $G$ represents in this case either the $F$ or the $H$  operations, but in general it could also represent a different group. 
 This property was already recognised in \cite{DurtNagler}, and abundantly used. This property is obvious in the Fourier case (addition and multiplication modulo $N$), but in the Hadamard case, we can explain it as follows:
 the Hadamard addition table respects parity. This is why in this case $\gamma_{H}=-1$.

 Actually, the structure of the Hadamard operation is elucidated if, formally, we express quartits as products of two qubits: 
$\ket{0}_{4}=\ket{0}_{2}\otimes\ket{0}_{2}$, $\ket{1}_{4}=\ket{0}_{2}\otimes
\ket{1}_{2}$, 
$\ket{2}_{4}=\ket{1}_{2}\otimes\ket{0}_{2}$, $\ket{3}_{4}=\ket{1}_{2}\otimes\ket{1}_{2}$. It is then easy to check the following property:
 If $\ket{i}_{4}=\ket{i_{1}}_{2}\otimes\ket{i_{2}}_{2}$, and $\ket{j}_{4}=
 \ket{j_{1}}_{2}\otimes\ket{j_{2}}_{2}$, 
 then $\ket{i\oplus_{H} j}_{4}=\ket{i_{1}\oplus_{F}j_{1}}_{2}\otimes\ket{i_{2}\oplus_{F}j_{2}}_{2}$.
   Remark that the property $\sum_{p=0}^{N-1} \gamma^{ (p\odot q)}=N\delta_{q,0}$ is true for the Fourier multiplication
 and the Hadamard multiplication as well. 
 
 It is instructive to look at the Fourier and the Hadamard duality transformations (they were intensively studied in Ref.\cite{DurtNagler}).
 
 \begin{table}
\begin{tabular}{c||c|c|c|c}
\hline $\gamma_F^{i\odot_{F} j}$  & $0$ & $1$ & $2$  & $3$\\
\hline \hline  0 & 1 & $1$ & $1$  & $1$ \\
1 & $1$ & $i$ & $-1$  & $-i$\\
2 & $1$ & $-1$ & $1$  & $-1$\\
3 & $1$ & $-i$ & $-1$  & $i$ \\
 \hline
\end{tabular}
\caption{The Fourier dual transformation in dimension 4. }\label{tab3F}
\end{table}

\begin{table}
\begin{tabular}{c||c|c|c|c}
\hline $\gamma_H^{i\odot_{H} j}$  & $0$ & $1$ & $2$  & $3$\\
\hline \hline  0 & 1 & $1$ & $1$  & $1$ \\
1 & $1$ & $-1$ & $1$  & $-1$\\
2 & $1$ & $1$ & $-1$  & $-1$\\
3 & $1$ & $-1$ & $-1$  & $1$ \\
 \hline
\end{tabular}
\caption{The Hadamard dual transformation in dimension 4. }\label{tab3H}
\end{table}

 We have also a dual relation
  between Bell states in both cases as shows the following generalization of the proof given in 
  Ref.\cite{DurtNagler} for the case $N=4$:
 
 \beqa
\ket{\tilde B_{m^*,n}}=N^{-1/2} \sum_{k=0}^{N-1} \gamma_{G}^{(k\odot_{G} n)}
\ket{\tilde k^*}\ket{\tilde{k\oplus_{G}m}}\\
=N^{-3/2} \sum_{k,l,j=0}^{N-1} \gamma_{G}^{(k\odot_{G} n)}\gamma_{G}^{\ominus_{G}(k\odot_{G} l)}
\gamma_{G}^{(k\oplus_{G}m)\odot_{G} j)} \ket{l^*}\ket{j}\\
=N^{-1/2} N\sum_{l,j=0}^{N-1}\delta_{n\ominus_{G}l\oplus_{G}j,0} 
\gamma_{G}^{(m\odot_{G} j)} \ket{l^*}\ket{j}\\
=N^{-1/2} \sum_{l=0}^{N-1}\gamma_{G}^{(m\odot_{G} j)} \ket{l^*}\ket{l\ominus_{G} n}=\ket{B_{\ominus_{G} n^*,m}}
\label{bellgeneral}
\eeqa

As it was shown in \cite{DurtNagler}, the dual states (and the generalized Bell states as well) can be derived from the knowledge of the fundamental
 permutation group $P_{G}$ defined as follows: 

\beq
P^i_{G}\ket{ j}=\ket{ j\oplus_{G} i}\label{Perm} \eeq

Obviously, this group possesses the following composition law: \beq
P^i_{G}P^j_{G}=P^{i \oplus_{G} j}_{G}\eeq
The dual states are invariant under this group:

\beqa P^i_{G}\ket{\tilde  j}=\frac{1}{\sqrt{N}}\sum_{k=0}^{N-1}\gamma^{(k\odot_{G} j)}\ket{k \oplus_{G} i}\\
=\frac{1}{\sqrt{N}}\sum_{k'=0}^{N-1}\gamma^{((k'\ominus_{G} i)\odot_{G} j)}\ket{k' }\\
=\gamma^{(\ominus_{G})( i\odot_{G} j) }\ket{\tilde j}\eeqa
This is also true for the generalized Bell states:
 
 \beqa P^i_{G}\ket{\tilde B_{m^*,n}}= P^i_{G}N^{-1/2}\sum_{k=0}^{N-1} \gamma_{G}^{(k\odot_{G} n)}
\ket{\tilde k^*}\ket{\tilde{k\oplus_{G} m}}\\
=N^{-1/2}\sum_{k=0}^{N-1} \gamma_{G}^{(k\odot_{G} n)}
\ket{\tilde k^*\oplus_{G} i^*}\ket{\tilde{k\oplus_{G} m \oplus_{G} i}}\\
=N^{-1/2}\sum_{k'=0}^{N-1} \gamma_{G}^{((k'\ominus_{G} i)\odot_{G} n)}
\ket{\tilde k'^*}\ket{\tilde{k'\oplus_{G} m }}\\ =  \gamma_{G}^{((\ominus_{G} i)\odot_{G} n)}\ket{\tilde B_{m^*,n}}
\eeqa

Note that the group $P^i_{G}$ acts covariantly on one of the qu$N$its and contravariantly onto the other one. 
It is easy to generalize the Fourier construction for arbitrary dimensions, and the Hadamard construction to dimension $2^m$, with $m$ a positive integer. 
Without much difficulty, it is also possible to generalize this construction to the dimension $p^m$, with $p$ a prime number. 
The Hilbert space is then seen as the product  ($m$ times) of $p$ dimensional Hilbert spaces, and the corresponding $\gamma_{G}$ is
 the $p$th root of unity $e^{i (2\pi  /p)}$ (by doing so, we obtain an extended parity, which takes $N$ values, and corresponds to the rest obtained after division by $p$). 
The addition is then obtained from the addition modulo $p$ restricted to each
 of the Hilbert spaces, similarly to the Hadamard construction in dimension 4 (componentwise addition of $m$-uples). 
 This is due to the fact that for dimensions equal to powers of prime, with the componentwise 
 addition derived from the addition modulo $p$ it is known that a finite field exists, 
 so there exists a multiplication which is distributive relatively to 
the addition (see \cite{Archer} and references therein for a recent contribution on the subject). It is  worth noting that a field structure is not necessary in order to derive
 a dual basis, and generalized Bell states as in this section:
 it could occur that ``exotic'' additions and multiplications exist for a given dimension $N$ 
 such that the multiplication is distributive and both operations are commutative (operations modulo $N$ are a good candidate). 
 Then a dual basis exists according to the construction described in this section. It could even be that the requirement of commutativity is too strong and could be relaxed, but this is out of the scope of the present paper.
\section{Construction of $N$-2 other dual bases}
The construction of other dual bases is straightforward, taking account of the fact that Bell states remain Bell states in the dual basis, thanks to the duality relation (\ref{bellgeneral}), and that they remain Bell states under the action of two groups:
 the 
 fundamental group of permutation $P_{G}$ expressed in the computational basis (\ref{Perm}),
  and the same group expressed in the dual basis $\tilde P_{G}$:
  \beq
\tilde P^i_{G}\ket{ \tilde j}=\ket{ \tilde {j\oplus_{G} i}}\label{permtilde} \eeq 
In the computational basis, these transformations are expressed as follows:

\beq
\tilde P^i_{G}\ket{ j}=\gamma^{j\odot_{G} i}\ket{  j}\label{permtildecompu} \eeq 
It is easy to check that $ P^i_{G}$ and $\tilde P^i_{G}$ form groups of unitary transformations with $N$ elements.
The Bell states are necessarily
   invariant under compositions of the operations (\ref{Perm}) and (\ref{permtilde}). In order to study the 
   group of transformations generated by the composition of 
  $ P^i_{G}$ and $\tilde P^i_{G} $, it appears to be convenient to introduce the following
   new and compact notation:

 \beq
 V^j_i=\tilde P^j_{G}. P^i_{G}\label{defV0};i,j:0...N-1 \eeq 
 Here, the product . expresses the matricial product (the usual composition law of two unitary transformations).

 As $V^0_0 (=\tilde P^0_{G}. P^0_{G}=\tilde P^0_{G}=P^0_{G}$) 
 is the identity, the following identities are fulfilled:
 
 \beq
\tilde P^i_{G}=V^i_0\label{defV1} \eeq 
\beq
 P^i_{G}=V^0_i\label{defV2} \eeq
 
  \beq
 V^j_i=V^j_0.V^0_i\label{defV3} \eeq 
 
 Note that $\tilde P^i_{G}$ and $ P^j_{G}$ do not commute: 
 \beqa \tilde P^j_{G}. P^i_{G}=
 \sum_{l=0}^{ N- l}\gamma^{l\odot_{G} j}\ket{  l}\bra{  l}.\sum_{k=0}^{ N- l}\ket{  k\oplus_{G} i}\bra{  k}\\
 = \sum_{k=0}^{ N- l}\gamma^{((k\oplus_{G} i)\odot_{G} j)}\ket{  k\oplus_{G} i}\bra{  k}\eeqa
 
 \beqa P^i_{G}. \tilde P^j_{G} =
 \sum_{k=0}^{ N- l}\ket{  k\oplus_{G} i}\bra{  k}.\sum_{l=0}^{ N- l}\gamma^{l\odot_{G} j}\ket{  l}\bra{  l}\\
 = \sum_{k=0}^{ N- l}\gamma^{(k\odot_{G} j)}\ket{  k\oplus_{G} i}\bra{  k}\\
 =\gamma^{\ominus_{G}(i\odot_{G} j)}\tilde P^j_{G}. P^i_{G}\eeqa
 Actually, we recover a commutation rule that is known as the Weyl commutation rule,
  and was already derived before in the study of mutually unbiased bases \cite{weyl, Englert}.
 By a straightforward computation, we can now derive the law of composition of these
 $N^2$ unitary transformations: 
 \beqa
 V^j_i.V^k_l=\tilde P^j_{G}. P^i_{G}.\tilde P^k_{G}. P^l_{G}\\
 =\gamma^{\ominus_{G}(i\odot_{G} k)}\tilde P^j_{G}.\tilde P^k_{G}. P^i_{G}. P^l_{G}\\
 = \gamma^{\ominus_{G}(i\odot_{G} k)} V^{j\oplus k}_{i\oplus l}\label{compo} \eeqa 

Up to a global phase, this looks like a groupal composition law. We shall consider the question of the phases soon. Note that the previous expression already contains in germ the quaternionic signature of the Pauli matrices, in the qubit case.
\subsection{The Fourier case (operations modulo $N$).}
If we compute directly the set of $N^2$ unitary operators $V^j_i$, with addition and multiplication modulo $N$, when $N$=2 and 3 (qubit and qutrit cases),
  we obtain respectively 3 subgroups with 2 elements and 4 subgroups with 3 elements (up to global phases). In general we expect thus that 
 the $N^2$ unitary operators $V^j_i$ form at least $N+1$ distinct sets, of $N-1$ elements, that (together with the identity) are closed under the composition law (up to global phases), and the identity (which corresponds to $N+1$ subgroups). 
 Moreover, on the basis of these examples, it is easy to check by direct computation that these $N+1$ subgroups form
  a repetitive pattern: each of them can be diagonalized in a certain orthogonal basis (that consists  of $N$ states that 
  we shall from now on label by an upper index $i$: 
  $1=\sum_{k}\ket{  e_{k}^i}\bra{  e_{k}^i}, k=0...N-1,i=0...N$). We know already the two first subgroups, they 
  correspond to the  the computational and the dual basis, and to the subgroups
   $\tilde P_{G}$ and $ P_{G}$ respectively. We find in dimensions 2 and 3 that, up to a global phase, the other subgroups can be parametrized as follows:  
  $V^{((i\ominus_{F} 1)\odot_{F} l)}_{l}=
  phase.\sum_{k} \gamma_{F}^{ k\odot_{F} l}\ket{  e_{k}^i}\bra{  e_{k}^i},
   l=0...N-1, i=2...N.$ 
   Let us assume (this will be proven later) that  for arbitrary dimension, there exist
    $N-1$ bases $\ket{  e_{k}^i}, k=0...N-1, i=2...N$ 
   such that $V^{((i\ominus_{F} 1)\odot_{F} l)}_{l}=\sum_{k} \gamma_{F}^{ k\odot_{F} l}\ket{  e_{k}^i}\bra{  e_{k}^i}$ up to a phase, and let us define the $U$ 
   operators as follows: \beq U^{(i\odot l)}_{l}=
  \sum_{k} \gamma_{F}^{ k\odot_{F} l}\ket{  e_{k}^i}\bra{  e_{k}^i}\label{defU}\eeq
  
  Obviously, the operators $U$ form, under composition a commutative group isomorph to $\tilde P_{F}$ and $ P_{F}$, and to the addition modulo $N$.
 We must still fix the phases. In the Fourier construction, this can be done if we note that  $U^{(i\odot_{F} l)}_{l}
 =(U^{(i\odot_{F} 1)}_{1})^l$ so that the phase relations between $U^{(i\odot_{F} l)}_{l}$ 
 and $V^{((i\ominus_{F} 1)\odot_{F} l)}_{l}$
  are fixed, in virtue of the law of composition \ref{compo}, provided we know the phase relation between 
  $U^{(i\odot_{F} 1)}_{1}$ and $V^{((i\ominus_{F} 1)\odot_{F} 1)}_{1}$. Indeed, iterating $l$ times the
   composition law \ref{compo}, we get that
    \beqa U^{(i\odot_{F} l)}_{l}=
  \sum_{k} \gamma^{ k\odot_{F} l}\ket{  e_{k}^i}\bra{  e_{k}^i}=(U^{(i\odot_{F} 1)}_{1})^l \nonumber\\
  =(V^{((i\ominus_{F} 1)\odot_{F} 1)}_{1})^l \cdot (U^{(i\odot_{F} 1)}_{1}/V^{((i\ominus_{F} 1)\odot 1}_{1})^l\nonumber\\
  =(U^{(i\odot_{F} 1)}_{1}/V^{((i\ominus_{F} 1)\odot_{F} 1}_{1})^l \cdot   \gamma_{F}^{-(i-1)l(l-1)/ 2}\cdot 
  V^{((i\ominus_{F} 1)\odot_{F} l)}_{l}\eeqa

 In order to fix the phase $U^{(i\odot_{F} 1)}_{1}/V^{((i\ominus_{F} 1)\odot 1}_{1}$, it is enough to impose that 
 $U^{(i\odot_{F} N)}_{N}=1$, so that we obtain the following relation:
 \beq U^{(i\odot_{F} l)}_{l}= \gamma_{F}^{-(i-1)l(l-1)/ 2}\gamma_{F}^{+(i-1)l(N-1)/ 2}V^{((i\ominus_{F} 1)\odot_{F} l)}_{l}\eeq
Note that this relation is also valid for $i=1$, which corresponds to the dual basis derived in the previous section. For $i=0$, we define the operators $U$ as follows: 
$U^{(0\odot_{F} l)}_{l}=U^{0}_{l}=V^{l}_{0}$, in agreement with the relation \ref{defU}.
One can check (we shall prove it later for an arbitrary group $G$) that the composition law for the $U$ operators is, as expected, the following: 
\beq U^{(i\odot_{F} l_{1})}_{l_{1}}\cdot U^{(i\odot_{F} l_{2})}_{l_{2}}=
U^{(i\odot_{F} (l_{1}\oplus_{F} l_{2}))}_{l_{1}\oplus_{F}  l_{2}}\eeq
Now that we know the exact expression for the operators $U$, we can derive easily the $N-2$ dual bases associated to the cyclic groups
 of length $N$ that consist of powers of $U^{(i\odot 1)}_{1}$ ($i\ominus_{F} 1=1...N-1$):
\beq \ket{  e_{k}^i}\bra{  e_{k}^i}={1\over N}\sum_{l} \gamma^{ \ominus_{F}k\odot_{F} l}U^{(i\odot_{F} l)}_{l}
\eeq
By a straightforward but lengthy computation that we do not reproduce here, we obtain that 
\beq Tr.\ket{  e_{k}^i}\braket{  e_{k}^i}{  e_{k}^j}\bra{  e_{k}^j}=\delta_{i,j} \eeq
which confirms our prime intuition about the existence of an orthonormal basis that diagonalizes the operators $U$ ($V$).

Actually, there is an elegant manner for reexpressing the operators $U$:
\beq U^{k\odot_{F}l}_{l}= \sum_{p=0}^{N-1} \ket{  e^{0,k}_{p\oplus_{F}l}}\bra{  e^{0,k}_{p}}\eeq
where the states $\ket{  e^{0,k}_{p}}$ are defined as follows:
\beq  \ket{  e^{0,k}_{p}}= \gamma_{F}^{  {(k-1)\over 2}.p.(p+N)} \ket{  e^{0}_{p}}\eeq
This shows that the basis that diagonalizes the operator $U^{k}_{l}$ is Fourier dual (according to the construction described in the first section) relatively to a basis that is, up to well-chosen phases, the computational basis. 
De facto, the $N-2$ new bases that we get by diagonalising the $U^{k.l}_l$ operators ($k=1...N$, $l=0...N-1$) 
are mutually unbiased relatively
 to the computational basis, because they are mutually unbiased relatively to the 
 computational basis, up to phases: 
 \beqa  \ket{  e^{k}_{i}}=  { 1\over \sqrt N}\sum_{i=0}^{  N-1}\gamma_{F}^{ - p. i} \ket{  e^{0,k}_{p}}\nonumber \\
 =  { 1\over \sqrt N}\sum_{i=0}^{  N-1}\gamma_{F}^{ - p. i} \gamma_{F}^{  {(k-1)\over 2}p(p+N)} 
 \ket{  e^{0}_{p}}\label{explicit}\eeqa
 
At first sight the expression of these bases is 
very close to an expression derived in \cite{Ivanovic} for odd prime dimensions (see footnote \ref{newG}). It is interesting to check whether these bases are also mutually unbiased relatively to the Fourier dual basis defined 
 in Eqn.\ref{fourierdual}. By a direct computation, we get that 
 \beqa Tr \ket{  e^{1}_{i}}\braket{  e^{1}_{i}}{  e^{k}_{j}}\ket{  e^{k}_{j}}
 =\nonumber \\
 \delta_{  k-1,0}\delta_{  i,j}+\nonumber \\
 (1-\delta_{  k-1,0}){  1\over N}\sum_{  l=0}^{  N-1}
 \delta_{  (k-1).l,0}\gamma^{- l\cdot(i-j-((1-k)(l+N)/2))}\eeqa
The factor $(1-\delta_{  k-1,0})\delta_{  (k-1).l,0}$ that appears in this expression is crucial: in prime dimensions, and in prime dimensions only, 
$(1-\delta_{  k-1,0})\delta_{  (k-1).l,0}=(1-\delta_{  k-1,0})\delta_{l,0}$, because there is no divider of 0 excepted 0 itself 
(the multiplication forms a division ring). Then
 $Tr \ket{  e^{1}_{i}}\braket{  e^{1}_{i}}{  e^{k}_{j}}\ket{  e^{k}_{j}}=1/N$ when $k\not= 1$, and it is easy to show by similar computations that the $N-2$ bases that we obtain by diagonalising the $U$
 operators are mutually unbiased relatively to the dual Fourier basis and also between themselves. 
 Otherwise, when the dimension $N$ is not a prime number, we obtain $N-1$ bases mutually unbiased relatively to
  the computational basis but not between themselves. Moreover, the $U$ operators are not in one to one relation with the $V$ operators, when the dimension is not prime. This is due to the fact that in our construction $U^{k.l}_{l}$ is proportional to $V^{(k-1).l}_{l}$, up to a phase.
   Now,  $V^{i}_{j}$ can always be expressed as $V^{(k-1).l}_{l}$, when $i$ and $j$ differ from 0, provided $j$ possesses a multiplicative inverse. Then, 
   $k=1+i/j$ and $l=j$. This is true only in prime dimensions. For instance, when $N=4$,
    there are $N+2$ subgroups
    of the $N^2$ $V$ operators, and not $N+1$ as expected and it is necessary to introduce a new cycle of length four, thus four new commuting $U$ operators (and so a new basis) in order to diagonalize all the $V$ operators.
   The subgroups are  
   $(V^{0}_{0},V^{1}_{0},V^{2}_{0}, V^{3}_{0})$,
    $(V^{0}_{0},V^{0}_{1},V^{0}_{2},V^{0}_{3})$,
    $(V^{0}_{0},V^{1}_{1},V^{2}_{2},V^{3}_{3})$,
    $(V^{0}_{0},V^{1}_{2},V^{2}_{0},V^{3}_{2})$,
     $(V^{0}_{0},V^{1}_{3},V^{2}_{2},V^{3}_{1})$, and $(V^{0}_{0},V^{2}_{1},V^{0}_{2},V^{2}_{3})$. The operators that belong to different subgroups are degenerated, this is why they can be diagonalized in different bases.
  Intuitively, we can understand the special role played by prime dimensions if we consider 
  the relation between bases that differ from the computational one:
  
  \beqa  \braket{  e^{l}_{j}}{  e^{k}_{i}}=  { 1\over  N}\sum_{p=0}^{  N-1}
  \gamma_{F}^{ p\odot (j\ominus i)} 
  \gamma_{F}^{{ (k\ominus l)\over 2}p(p+N)} \label{in-product}\eeqa

  This suggests some cyclicity between the $N$ dual bases $e^{l}_{j} (l:1...N, j: 0...N-1)$, and, effectively, in prime dimensions a new symmetry appears, for which we did not find a better
   name than group-relativity. What does it consist of? 
  Now that we have at our disposal an explicit expression (\ref{explicit}) for these dual bases,
   we can evaluate the unitary  transformation $W^k$ that maps the computational basis onto the $k$th basis:
  \beq  W^k=  \sum_{i=0}^{  N-1}\ket{  e^{k}_{i}}\bra{  e^{0}_{i}}\label{W}\eeq
  We can thus reevaluate the situation from the point of view of the $k$th basis. For instance, we get that, up to phases, and up to a bijective redistribution of their $N^2$ labels, 
  the error operators $V^m_{n}$ are still error operators from the point of view of the $k$th basis: 
  $(W^k)^+V^m_{n}W^k=phase.V^n_{(k-1).n-m}$. This relation is true in arbitrary dimension, but it is only in prime dimension
   that the redistribution of the labels is bijective. It is easy to explain this relation if we derive the $V$ operators from the 
   generalized Bell states defined in Eqn.\ref{bell} as we shall do now, according to a standard
    procedure in quantum cryptography \cite{CERFPRL,boure1}. This derivation also explains why the $V$ operators can be viewed as error operators. Let us assume that Alice and Bob share the maximally entangled state
    $\ket{B_{0^*,0}}_{AB}=N^{-1/2} \sum_{k=0}^{N-1} \ket{k^*}_{A}\ket{k}_{B}$ and that Alice measures the state $\ket{k^*}_{A}$,
     so that Bob gets the collapsed state $\ket{k}_{B}$. This is a way for Alice to transmit qu$N$its to Bob. 
     Let us assume now that an eavesdropper Eve controls the source and replaces 
     $\ket{B_{0^*,0}}_{AB}$ by $\ket{B_{n^*,m}}_{AB}$. When Alice projects her qu$N$it onto $\ket{k^*}_{A}$, then Bob
      receives, instead of $\ket{k}_{B}$, $V^m_{n}\ket{k}_{B}$ (up to a global, irrelevant, phase). Beside,
       the transformation \ref{explicit} between the computational basis and the $k$th basis can be decomposed into the composition of a phase transformation that 
      maps  $\ket{  e^{0}_{p}}$ onto $\ket{  e^{0,k}_{p}}=\gamma_{F}^{  {(k-1)\over 2}p(p+N)}$, followed by a dual 
      transformation similar to the transformation described in Eqn.\ref{fourierdual}. It is easy to check, by direct computation, that 
      the first transformation sends the Bell state $\ket{B_{n^*,m}}$ onto $\ket{B_{n^*,m-n(k-1)}}$ (up to a phase),
       and that the dual transformation sends, according to the Eqn.\ref{bellinv}, $\ket{B_{n^*,m-n(k-1)}}$ onto 
       $\ket{B_{-m+n(k-1)^*,n}}$ (up to a phase). 
       Accordingly, the $V$ and $U$ operators are bijectively intertwined when one passes from the computational basis to
        any of the $N$ other bases that diagonalize these operators; for instance, $V^m_{n}$ is mapped onto $V^n_{(k-1).n-m}$, up to a phase. This mapping is bijective only in prime dimensions. 
      This explains the special role played by prime dimensions (or more generally by the existence of a field): the $N+1$ bases are, roughly speaking, treated on the same footing. This also explains our empirical remark about the repetitive nature of the pattern obtained in prime dimensions 2 and 3 (at the beginning of this section).
   
   Note that, by construction, we are free to redefine the labels of the basis states, up to arbitrary shifts. In prime dimensions we also are
    free to redefine these labels up to arbitrary dilations. We can also adopt for the value of $\gamma_{F}$ any $N$th root
   of unity 
 ($\gamma^j_{F} =e^{ i .j2\pi/N}, j=1...N-1$), without losing any of the properties that 
 we obtained when we chose $j=1$ (the powers of such numbers form a group isomorph to the addition modulo $N$, in prime dimensions only; for instance -1, which is also the fourth root of unity generates a cycle of length 2 but not of length 4).
  If we redefine the square root of unity 
 the phase factors $\gamma_{F}^{  {(k-l)\over 2}p(p+N)}, k-l=1...N-1 $ become
  $\gamma_{F}^{  {j(k-l)\over 2}p(p+N)}=\gamma_{F}^{  {(k'-l)\over 2}p(p+N)}
  k-l, k'-l=1...N-1, k'-l=j\odot_{F}(k-l). $ 
  This shows that by adopting a new determination of the $N$th root of unity, we simply relabellize the 
 $N-1$ dual bases obtained in our approach. 
 
  In conclusion, when 
  the underlying algebraic structure is a finite field, we get an extremely symmetric (``group-relativistic'') pattern: 
  roughly speaking no basis and no basis state are privilegged.
\subsection{The Hadamard case.}
It is worth mentioning that similar properties are valid in the Hadamard case. Then, we fix the phases by requiring that the square of the operators $U$ is the identity, which imposes that
 \beq U^{(i\odot_{H} l)}_{l}= \gamma_{H}^{{1 \over 2 }\cdot(i-1)\odot_{H}l\odot_{H}l}
 V^{((i- 1)\odot_{H} l)}_{l}\eeq
This is sufficient in order to obtain the right composition law for the operators $U$:
\beqa U^{(i\odot_{H} l_{1})}_{ l_{1}}.U^{(i\odot_{H}  l_{2})}_{ l_{2}}= \nonumber \\
\gamma_{H}^{{1 \over 2 }\cdot(i-1)\odot_{H}l_{1}\odot_{H}l_{1}}\gamma_{H}^{{1 \over 2 }\cdot (i-1)
\odot_{H}l_{2}\odot_{H}l_{2}}
 V^{((i-1)\odot_{H} l_{1})}_{l_{1}}V^{((i-1)\odot_{H} l_{2})}_{l_{2}}=\nonumber \\ 
 \gamma_{H}^{{1 \over 2 }\cdot(i-1)\odot_{H}((l_{1}\odot_{H}l_{1})\oplus_{H}   (l_{2}\odot_{H}l_{2}) \ominus_{H}
   2.(l_{1}\odot_{H}l_{2})}V^{((i- 1)\odot_{H} (l_{1}\oplus l_{2}))}_{l_{1}\oplus l_{2}}\nonumber \\
=\gamma_{H}^{{1 \over 2 }\cdot(i-1)\odot_{H}(l_{1}\oplus_{H}l_{2})^2}
 V^{((i-1)\odot_{H} (l_{1}\oplus l_{2}))}_{l_{1}\oplus l_{2}}\nonumber \\
   =U^{(i\odot_{H} (l_{1}\oplus l_{2}))}_{l_{1}\oplus l_{2}}\eeqa
 By direct computation, we find 5 (4+1) subgroups of 4 elements (3+ the identity) 
 that are listed in the table 5, and are in agreement with the litterature on the subject. The three last lines are obtained from the products of the transformations that belong to the two first lines ($\tilde P_{i}$ and $P_{j}$).  
 They are expressed as products of two qubit operators, for reasons of simplicity. These subgroups can be shown to be diagonal in 5 mutually
  unbiased bases  $\ket{  e^{k}_{p}}$ (k=0,1,..5, p=0...3) that we do not reproduce here. Actually, it is possible in the
   double Hadamard case considered here ($N$ = 4), but also in general, to obtain an explicit expression of all the mutually unbiased bases (when the dimension is a prime power) that is unambiguously determined by the tables of multiplication and addition of the underlying field as we shall show now.
  \begin{table}
\begin{tabular}{c||c|}
\hline $U^{i\odot j}_{j}$  & \\
\hline \hline i= 0 & 1 , $\sigma_{Z1}\otimes 1_{2}$ , $1_{1}\otimes \sigma_{Z2}$ , $\sigma_{Z1}\otimes \sigma_{Z2}$ \\
i=1 & $1$ , $1_{1}\otimes \sigma_{X2}$ , $\sigma_{X1}\otimes 1_{2}$  , $\sigma_{X1}\otimes \sigma_{X2}$\\
  & $1$ , $\sigma_{Y1}\otimes 1_{2}$ , $1_{1}\otimes \sigma_{Y2}$ , $\sigma_{Y1}\otimes \sigma_{Y2}$\\
 & $1$ , $\sigma_{Z1}\otimes \sigma_{X2}$ , $\sigma_{X1}\otimes \sigma_{Y2}$  , $\sigma_{Y1}\otimes \sigma_{Z2}$\\
 & $1$ , $\sigma_{Y1}\otimes \sigma_{X2}$ , $\sigma_{Z1}\otimes \sigma_{Y2}$  , $\sigma_{X1}\otimes \sigma_{Z2}$\\
 \hline
\end{tabular}
\caption{The Hadamard subgroups in dimension 4. }\label{tab1}
\end{table}
\subsection{General expression of the mutually unbiased bases when the dimension is a prime power.}
Let us assume that the dimension is a prime power $N=p^m$, so that there exists a field $G$ with $N$ elements, and two operations $\odot_{G}$ and $\oplus_{G}$.
Let us define the $U$ operators as follows:
\beq U^{k\odot_{G}l}_{l}= \sum_{q=0}^{N-1} \ket{  e^{0,k}_{q\oplus_{G}l}}\bra{  e^{0,k}_{q}};l:0...N-1, k:1...N\eeq
where $ \gamma_{G}=e^{i.2\pi \over p}$ and the states $\ket{  e^{0,k}_{q}}$ are defined as follows:
\beq  \ket{  e^{0,k}_{q}}= (\gamma_{G}^{  (k-1)\odot_{G} q\odot_{G}q })^{1\over 2} \ket{  e^{0}_{q}}q:0...N-1\eeq
Note that in this hybrid expression certain operations are expressed in terms of the usual operations (the corresponding field is the set of complex numbers). 
In virtue of the identities $\gamma_G^{i}\gamma_G^{j}=\gamma^{(i\oplus_G j)}$, 
 and $\sum_{p=0}^{N-1} \gamma^{ (p\odot_G q)}=N\delta_{q,0}$, the   ($N^2-N$) $U$ operators are diagonal 
 in the $N$ dual bases defined by a generalisation of the expression \ref{explicit}:
 \beqa  \ket{  e^{k}_{i}}=  { 1\over \sqrt N}\sum_{i=0}^{  N-1}\gamma_{G}^{ \ominus_{G} q\odot_{G} i} \ket{  e^{0,k}_{q}}\nonumber \\
 =  { 1\over \sqrt N}\sum_{i=0}^{  N-1}\gamma_{G}^{ \ominus_{G} q_{G} i} (\gamma_{G}^{  (k-1)\odot_{G} q\odot_{G}q })^{1\over 2}
 \ket{  e^{0}_{q}}\label{genexplicit}\eeqa
  so that the relation \ref{defU} is still valid:
 \beq U^{(i\odot_{G} l)}_{l}=
  \sum_{k} \gamma_{G}^{ k\odot_{G} l}\ket{  e_{k}^i}\bra{  e_{k}^i}\label{defUgen}\eeq
 Let us now check by direct computation that the $N$ bases obtained so are mutually unbiased. 
  \beq  \braket{  e^{k}_{i}}{  e^{l}_{j}}= 
  { 1\over N}\sum_{q=0}^{  N-1}\gamma_{G}^{ \ominus_{G} q\odot_{G} (i\ominus_{G}j)}
  (\gamma_{G}^{  ((l-1)\ominus_{G}(k-1))\odot_{G} q\odot_{G}q })^{1\over 2} 
  \nonumber \eeq
  \beqa  \braket{  e^{k}_{i}}{  e^{l}_{j}}.\braket{  e^{l}_{j}}{  e^{k}_{i}} = \nonumber \\
    { 1\over N^2}(\sum_{q=0}^{  N-1}\gamma_{G}^{ \ominus_{G} q\odot_{G} (i\ominus_{G}j)}
  (\gamma_{G}^{  ((l-1)\ominus_{G}(k-1))\odot_{G} q\odot_{G}q })^{1\over 2}).
  (\sum_{q'=0}^{  N-1}\gamma_{G}^{ \ominus_{G} q'\odot_{G} (j\ominus_{G}i)}
  (\gamma_{G}^{  ((k-1)\ominus_{G}(l-1))\odot_{G} q'\odot_{G}q' })^{1\over 2}) \nonumber \\
  = { 1\over N^2}(\sum_{q=0}^{  N-1}\gamma_{G}^{ \ominus_{G} q\odot_{G} (i\ominus_{G}j)}
  (\gamma_{G}^{  ((l-1)\ominus_{G}(k-1))\odot_{G} q\odot_{G}q })^{1\over 2}).
  (\sum_{t=0}^{  N-1}\gamma_{G}^{ \ominus_{G} (q\oplus_{G}t)
  \odot_{G} (j\ominus_{G}i)}
  (\gamma_{G}^{  ((k-1)\ominus_{G}(l-1))\odot_{G}  (q\oplus_{G}t)\odot_{G} (q\oplus_{G}t) })^{1\over 2}) \nonumber \\
  = { 1\over N^2}(\sum_{q,t=0}^{  N-1}\gamma_{G}^{ t\odot_{G} (j\ominus_{G}i)}
  (\gamma_{G}^{ 2. ((k-1)\ominus_{G}(l-1))\odot_{G} q\odot_{G}t })^{1\over 2}.
  (\gamma_{G}^{  ((k-1)\ominus_{G}(l-1))\odot_{G}  t\odot_{G} t })^{1\over 2}) 
  \nonumber \\
  = { 1\over N^2}(\sum_{q,t=0}^{  N-1}\gamma_{G}^{ ((t\odot_{G} (j\ominus_{G}i)\oplus_{G}  ((k-1)\ominus_{G}(l-1))
  \odot_{G} q\odot_{G}t) }.
  (\gamma_{G}^{  ((k-1)\ominus_{G}(l-1))\odot_{G}  t\odot_{G} t })^{1\over 2}) \nonumber \\
  = { 1\over N^2}(\sum_{t=0}^{  N-1}(\sum_{q=0}^{  N-1}
  \gamma_{G}^{ ( ((k-1)\ominus_{G}(l-1))\odot_{G}t\odot_{G} q) }).\gamma_{G}^{ (t\odot_{G} (j\ominus_{G}i)) }
  .(\gamma_{G}^{  ((k-1)\ominus_{G}(l-1))\odot_{G}  t\odot_{G} t })^{1\over 2})\nonumber \\
  = { 1\over N}\sum_{t=0}^{  N-1}\delta_{((k-1)\ominus_{G}(l-1))\odot_{G}t,0)}
.\gamma_{G}^{ (t\odot_{G} (j\ominus_{G}i)) }
  .(\gamma_{G}^{  ((k-1)\ominus_{G}(l-1))\odot_{G}  t\odot_{G} t })^{1\over 2} \nonumber \\
 =\delta_{  (k-1)\ominus_{G}(l-1),0}\delta_{  i,j}+
 (1-\delta_{ (k-1)\ominus_{G}(l-1),0}){  1\over N}\sum_{  l=0}^{  N-1}
 \delta_{  t,0}.\gamma_{G}^{ (t\odot_{G} (j\ominus_{G}i)) }
  .(\gamma_{G}^{  ((k-1)\ominus_{G}(l-1))\odot_{G}  t\odot_{G} t })^{1\over 2}\nonumber \\
  =\delta_{  k,l}\delta_{  i,j}+(1/N). (1-\delta_{ k,l})\eeqa
We made use of the fact that there is no divider of 0 excepted 0 itself 
(the multiplication $\odot_{G}$ forms a division ring) \footnote{ \label{newG}Curiously, the expression
 for the states of the mutually unbiased bases is at first
 sight
 close to the solution of ref. \cite{Ivanovic} when the dimension is an odd prime,
  but this is true only at first sight. 
  To our knowledge, the hybrid expression with the $G$ products 
  instead of the usual products (and with the power 1/2) that we derived in this paper is new (and more general because it works in dimensions prime odd AND even, and in their powers ). 
  If this is not the case, we would be grateful to the reader who would inform us about the relevant reference. We were informed however that the approach through the $V$ operators is already known and that they form what is called the Pauli group (Refs. \cite{Rubin},  \cite{Yu} and \cite{india})}.
 For $i=0$, we define the operators $U$ as follows: 
$U^{(0\odot_{G} l)}_{l}=U^{0}_{l}=V^{l}_{0}=\sum_{k=0}^{ N- l}
\gamma^{(k\odot_{G} l)}\ket{  e_{k}^0}\bra{  e_{k}^0}$, in agreement with the relation \ref{defUgen}. These operators 
are diagonal in the computational basis which, obviously, is mutually unbiased relatively to the $N$ other bases.
  
  Beside, the $N^2$ $U$ operators are equal, up to a phase, to the $V$ operators which are defined as usually by the relation 
  $ V^j_i=\sum_{k=0}^{ N- l}\gamma^{((k\oplus_{G} i)\odot_{G} j)}\ket{ e_{ k\oplus_{G} i}^0}\bra{  e_{k}^0}$.
 The phase relation is the following:
  \beq U^{(i\odot_{G} l)}_{l}= (\gamma_{G}^{\ominus(i-1)\odot_{G}l\odot_{G}l})^{1\over 2}V^{((i -1)\odot_{G} l)}_{l}\eeq
  Note that it differs slightly from the phase relation introduced
   in the Fourier case, without remarkable consequence,
   excepted a gain in simplicity in the present case. Actually, we recover the Fourier case by
    posing formally $N$ =0 in the expression 
   \ref{explicit}, the Hadamard phase convention by replacing $-1$ by $(-1)^{-1}$, and
    the Ivanovic solution for odd primes value of $N$ \cite{Wootters} (they differ by a power 2, but 2 is not a divider of 0 in odd prime dimensions, so that the solutions are equivalent, up to a bijective relabelling of the bases). Note that as the Bell states are defined as a product of contra-covariant states,
     many phases puzzles are automatically
     solved in their case (by the way let us apologize for several inconsistencies
      in the sign of the phase for the dual bases, we shall correct them in a next version
       of the paper). It is our belief anyhow that there is no ``good'' phase convention.

The $U$ operators form a group, whenever they admit the same eigenbasis, in accordance with the relation
 \ref{defUgen} as we shall check now.
 \beqa U^{(i_{1}\odot_{G} l_{1})}_{l_{1}}.U^{(i_{2}\odot_{G} l_{2})}_{l_{2}}= (\gamma_{G}^{\ominus(i_{1}-1)\odot_{G}l_{1}\odot_{G}l_{1}})^{1\over 2} (\gamma_{G}^{\ominus(i_{2}-1)\odot_{G}l_{2}\odot_{G}l_{2}})^{1\over 2}
 V^{((i_{1} -1)\odot_{G} l_{1})}_{l_{1}}.V^{((i_{2} -1)\odot_{G} l_{2})}_{l_{2}}=\nonumber \\
 (\gamma_{G}^{\ominus(i_{1}-1)\odot_{G}l_{1}\odot_{G}l_{1}})^{1\over 2}  
 (\gamma_{G}^{\ominus(i_{2}-1)\odot_{G}l_{2}\odot_{G}l_{2}})^{1\over 2}
 (\gamma_{G}^{(l_{1}\odot\ominus(i_{2}-1)\odot_{G}l_{2}) \oplus  (l_{1}\odot\ominus(i_{2}-1)
 \odot_{G}l_{2}}))^{1\over 2}
 V^{((i_{1} -1)\odot_{G} l_{1}\oplus (i_{2} -1)\odot_{G} l_{2})}_{l_{1}\oplus l_{2}}=\nonumber \\
 (\gamma_{G}^{\ominus(i_{1}-1)\odot_{G}l_{1}\odot_{G}l_{1}})^{1\over 2}  
 (\gamma_{G}^{\ominus(i_{2}-1)\odot_{G}l_{2}\odot_{G}l_{2}})^{1\over 2}
 (\gamma_{G}^{(l_{1}\odot\ominus(i_{2}-1)\odot_{G}l_{2}) \oplus  (l_{1}\odot\ominus(i_{2}-1)
 \odot_{G}l_{2}}))^{1\over 2}
 V^{((i_{1} -1)\odot_{G} l_{1}\oplus (i_{2} -1)\odot_{G} l_{2})}_{l_{1}\oplus l_{2}}\nonumber \eeqa
 Henceforth, when $(i_{1} -1)=(i_{2} -1)=(i -1)$, we obtain:
 \beq U^{(i\odot_{G} l_{1})}_{l_{1}}.U^{(i\odot_{G} l_{2})}_{l_{2}}=U^{(i\odot_{G}( l_{1}+l_{2}))}_{l_{1}+l_{2}} \eeq
\section{Quantum tomography.}
It is easy to check that the simple Hadamard case and the Fourier case for $N=2$ are equivalent, and that we recover
 the well-known Pauli operators. Our approach provides thus a simple way to generalize these operators to arbitrary dimensions,
  although we obtain a maximal set of mutually unbiased bases only for certain dimensions. 
  Anyhow, the set of bases that we obtain, for instance through the Fourier approach, in arbitrary dimension,
   makes it possible to derive a set of at least $N+1$ bases, mutually unbiased relatively to the computational basis,
    which allow us to make a complete tomography of an arbitrary quantum state. 
  For instance, it is easy to show that the following equivalence is fulfilled: 
 \beq\ket{  e_{i\oplus j}^0}\bra{  e_{i}^0} =(1/N)\sum_{k,l}V^{k}_{l}Tr.((V^{k}_{l})^+.\ket{  e_{i\oplus j}^0}\bra{  e_{i}^0})\eeq
Indeed, $(V^{k}_{l})^+=\sum_{m} \gamma^{\ominus(m\oplus l)\odot k}\ket{  e_{m}^0}\bra{  e_{m\oplus l}^0}$ so that 
$Tr.(V^{k}_{l})^+\ket{  e_{i\oplus j}^0}\bra{  e_{i}^0}$=$\gamma^{\ominus(i\oplus l)\odot k}\delta_{l,j}$ and 
$(1/N)\sum_{k,l}V^{k}_{l}\gamma^{\ominus(i\oplus l)\odot k}\delta_{l,j}$ =
$(1/N)\sum_{k}V^{k}_{j}\gamma^{\ominus(i\oplus j)\odot k}$

=$(1/N)\sum_{k}\sum_{m} \gamma^{((m\oplus j)\ominus(i\oplus j))\odot k}\ket{  e_{m\oplus j}^0}\bra{  e_{m}^0}$

=$(1/N)\sum_{k}\sum_{m} N.\delta_{m,i}\ket{  e_{m\oplus j}^0}\bra{  e_{m}^0}=\ket{  e_{i\oplus j}^0}
\bra{  e_{i}^0}.$
By linearity, we obtain a similar development for any linear operator $L$: 
$L=(1/N)\sum_{k,l}V^{k}_{l}Tr.((V^{k}_{l})^+.L)$. Formally, if we express the density matrix as a $N^2$ dimensional state,
 the previous identity is equivalent to the 
fact that the Bell states form an orthonormal basis. The same development is valid
 for what concerns the $U$ operators,
 provided we defined as many subgroups of $N$ $U$ operators as there are bases that diagonalize the $V$ operators
  (there are certainly $N+1+i$ of them with $i$ positive, and equal to zero for prime dimensions only).
We can always define the $U$ operators in such a way that they are equal to the $V$ operators up to a phase, then they allow us to perform tomography:
$L=(1/N)\sum_{k,l}U^{k}_{l}Tr.((U^{k}_{l})^+.L)$.

In particular, when $L$ represents an arbitrary density matrix, we can obtain a complete knowledge
 about it by measuring the $N^2-1$ operators $V^k_{l}k,l\not= 0$,
 or because they are in one to one correspondence by measuring in the $N+1+i$ bases that diagonalize the $V$ operators 
 the $(N-1)(N+1+i)$ average values 
  $\bra{  e_{j}^k}L\ket{  e_{j}^k}, k:0...N+1+i, j\not=0$
   (the remaining values $\bra{  e_{0}^k}L\ket{  e_{0}^k}, k:0...N+1+i$ being obtained by 
   normalisation) . 
   
   Our treatment of the Fourier case, valid in arbitrary dimensions, confirms that, according
   to the reference \cite{Ivanovic}, if we want to perform tomography of
    a quantum state with a minimal number of bases ($N+1$), it
    is necessary that these
   $N+1$ bases are 
  mutually unbiased (informationnally independent). Indeed, when the dimension is not prime, the data
   collected in one basis allow us to gather
     by inference some information about the data collected in other bases so that these data are not independent.
      In order to evaluate the density matrix, we must evaluate $N^2-1$ independent positive parameters,
       so that when our data are not independent we need more than $N+1$ bases ($N+1+i$) in order to collect
        all the necessary information. Of course, tomography is possible in arbitrary dimensions, and it is not obligatory
         to measure the state in mutually unbiased bases. The
       procedure  is simply a bit easier and cheaper in prime dimension.

   \section{Open questions.}
       \subsection{Back to quantum information.}
    If we consider seriously the option according to which the symmetries studied here tell us something deep about the quantum world,
     about quantum information and so on, it is worth mentioning two interesting properties of the Bell states.
   
   Roughly speaking, the first property is the following. Let us consider a system that consists of two entangled $N^2$ dimensional systems
    that are prepared in a pure state. Formally, we can express the bases that biorthogonalize the full state as Bell states. 
    We introduce so a formal Alice and a formal Bob for the first system, a formal Eve and a formal Eve' for the second one (yes). Let
     us assume that this state is invariant under the label shift that corresponds to the addition $\oplus_{G}$ that
     we introduced in the paper. Then, it is easy to prove, by a direct generalisation of the proof given in ref.\cite{DurtKwek} that the following property is valid:
     
     (1 A) the full state obeys a particular, three parameters combination of products of Bell states; (1 B) this superposition can be decomposed
      into families that do not interfere. Everything happens thus as if, from the point of view of Eve, different families of (products of) Bell
states were separated by a classical super-selection rule. Each family constitutes thus a maximal block of information, and no gain can be expected from quantum coherence.
      
       We must confess that we do not fully understand the meaning of this property. 
     Anyhow it is useful when we want to optimize Eve's information, an old problem of quantum cryptography. 
    
Indeed, it can be shown \cite{DurtKwek} that, as a consequence, Eve will maximize her information by choosing a cloning state that obeys Cerf's 
ansatz \cite{CERFPRL}:
\beq \label{CERF}\ket{\Psi}_{ABEE^\prime}=
\sum_{m,n=1}^{N-1}a_{m,n}\ket{B^{\psi}_{m^*,n}}_{AB}
\ket{B^{\psi}_{m,-n^*}}_{EE^\prime}\eeq
This state is biorthogonal in the Bell bases  $\ket{B^{\psi}_{m^*,n}}_{AB}$ and
$\ket{B^{\psi}_{m^*,n}}_{EE'}$. The second property is an interesting symmetry that was derived by N. Cerf for the modulo $N$ operations,
 and according to which such states are also biorthogonal in the Bell bases $\ket{B^{\psi}_{m^*,n}}_{AE}$ and
$\ket{B^{\psi}_{m^*,n}}_{BE'}$. This fact is based on the following identity, a Fourier like relation:

\beq\braket{B^G_{m,n A,B} \otimes B^G_{m,n E,E'}}{B^G_{i,j A,E}\otimes B^G_{i,j B,E'}}={1\over  N}\gamma_{G}^{i\odot_{G} n} .\gamma^{ m\odot_{G} j}\eeq

 We proved it for the Galois field with 4 elements in the ref.\cite{DurtNagler} (without making use of the multiplication at the time, and without knowing that it was a field that we were playing with) and it is easy
 to convince oneself that the proof is valid for general structures $G$. This property is important in cryptography,
  because it is a duality relation from which it is possible to derive a general trade-off
 relation between Eve and Bob's information \cite{CERFPRL}, but we are convinced that it also reveals a deeper complementarity.
  This complementarity is the following: when the entanglement of a system $X$ with a system $Y$ increases, the purity of the reduced state of $X$
 obtained by tracing out $Y$'s degrees of freedom, decreases. This duality property provides a new, modern, interpretation of the Bohrian complementarity
  between unbiased observables \cite{N=10}. For instance, the erasure of coherence that appears when the which-way information increases, in two-slits like experiments, could be explained in terms of entanglement only, and not by invoking an hypothetic stochastic disturbance
   that would be brought by the (supposedly classical) measurement process. To our knowledge, this complementarity has not yet been studied
    from a ``Bell state'' perspective. It is out of the scope of this paper to attack the problem, but it is worth mentioning that all the tools are present. 
    For instance, if we want to study the trade off between the purity
 of the reduced density matrix of the $X$ system (Alice and Bob) versus the entanglement of Alice with Eve (or the information of Eve) , it is sufficient to observe that this purity
  is a highly non-linear function of the squared moduli of the $a_{m,n}$ elements, while the entanglement with Eve (or the information of Eve) is a highly
   non-linear function of the squared moduli of Fourier transforms of the $a_{m,n}$ elements. It is also worth noting that in the case under study
    the mutual informations of Eve relatively to Alice's signal and Bob's signal symmetrically coincide.

    There remains another puzzling, open, question: we wrote in the second section that, roughly speaking, all bases are treated
   on the same footing. This is not absolutely true. Obviously the computational basis plays a privilegged role in our approach, and the dual basis too, to a lesser extent. It would be worth finding the larger structure (if it exists)
   that would be (this is just a vague intuition at this level) self-similar,
    when one repeats all the previously described construction, leaving from one arbitrary basis.
     It might well be that the different determinations of the $N$th root of unity play a role in this new, larger structure. It could also occur that it has something to do with the square of a field with $N$ elements (thus a field with $N^2$ elements), and with a group of permutations of $N^2$ elements, the indices of the Bell states.      Anyhow, it is certain that there is much
     more to say about the kind of structures that we sketched in the present work. It could be that the answer comes from the theory of algebraic field extensions \cite{Rubin}.

        \section{Conclusions and comments.}
 We obtained thus a generalization, valid in arbitrary dimension, of the Pauli operators (see also Refs. \cite{Rubin},  \cite{Yu} and \cite{india} for previous similar results). We also rederived in a straightforward manner the well-known error operators
    that found numerous applications in quantum cryptography \cite{CERFPRL,boure1}, and in quantum information theory \cite{nielsen}. 
   This construction is valid whenever certain ingredients are present:
    existence of two commutative relations between $N$ elements: an addition (that forms a group) and a multiplication, 
    distributive relatively to the addition, but also existence of a representation of 
    these groups by a complex phase $\gamma_{G}$, such that the following properties are true:
 $\gamma_G^{i\odot_G j}\gamma_G^{i\odot_G k}=\gamma^{i\odot_G (j \oplus_G k)}$, 
 $\sum_{p=0}^{N-1} \gamma^{ (p\odot_G q)}=N\delta_{q,0}$ and so on. 
 It is out of the scope of the present paper to study in detail all the structures that are 
 present here. For instance it would be worth investigating in depth the role played by 
 fields in this context and the stronger symmetries that appear when the multiplication
  is a division ring, and also understanding better the 
  interrelation between fields and mutually unbiased bases (what we did but maybe not with all the requested degree of generality).
   Note that we checked by direct computation that the bases that we obtained in the Fourier construction for dimensions 2 and 3, and in the Hadamard construction for dimensions 2 and 4 coincide with the results that can be found in the litterature.
  It is easy to show that in odd prime dimensions we recover Ivanovic's solution \cite{Ivanovic}. We expect a full agreement for arbitrary dimensions when these dimensions are powers of primes (odd or even) \cite{Wootters,Zeil}. 
  It is out of the scope of the present paper to compare our results with the resuts of refs.\cite{Wootters,Rubin} when the dimensions are powers of primes 
  ($m$ larger than 1, $N$ different from 4), because the corresponding tools are somewhat complex and sophisticated. 
  
  As we mentioned in the previous section, the Bell state approach makes it possible to attack
   the question of security of quantum cryptography,
   and also deeper questions related to the complementarity principle \cite{N=10}. Maybe that it also learns us something fundamental 
   about the nature of information \cite{Qinfo}. Certainly, this chapter is not closed yet. Note that we were able recently, thanks to the new tools derived in this paper, to derive an elegant solution for the mean king's problem \cite{vaid,Englert,2003,arXiv}.

  Finally, we are curious to learn more about the extremely symmetric pattern that we guessed, during the completion of this work.

     It is on these questions that we leave the reader, and with a personal remark for which we shall use the ``I'' form instead of the academic ``we'' form:
     
    {\it During the completion of this work, I was simply happy
      to rediscover by myself an intrinsic symmetry of finite dimensional Hilbert spaces,
       simple and beautiful.
       I realized afterwards by checking the references that maybe I increased
        a bit the simplicity of the description. I hope that I was able to communicate to
        the reader a part of my aesthetic pleasure, and of my amazement
         for these elegant and, 
        according to me, fundamental symmetries.}

    \leftline{\large \bf Acknowledgment}
\medskip The author acknowledges
a Postdoctoral Fellowship of the Fonds voor Wetenschappelijke Onderzoek,
Vlaanderen and also support from the
IUAP programme of the Belgian government, and the grant V-18.  
Many thanks to Prof.A. Rimini for his kind hospitality at the university of Pavia where this 
work was completed. Sincere thanks to Prof. M. D'Ariano and Dr. M. Sacchi for discussions and comments. 
 
 {\it More personnally, I devote this work to the memory of my
  regreted friend Mr Do Ba Vinh, who learnt me that civilisation is not the privilegge 
  of any nation.}

\end{document}